# VIBES: A Multi-Scale Modeling Approach Integrating Within-Host and Between-Hosts Dynamics in Epidemics

Paulo C. Ventura[1], Yong Dam Jeong[2,3], Maria Litvinova[4], Allisandra G. Kummer[1], Shingo Iwami[2], Hongjie Yu[5,6,7], Stefano Merler[8], Alessandro Vespignani[9], Keisuke Ejima[10], Marco Ajelli[1,*]

[1]Laboratory for Computational Epidemiology and Public Health, Department of Epidemiology and Biostatistics, Indiana University School of Public Health, Bloomington, IN, USA

[2]Interdisciplinary Biology Laboratory (iBLab), Division of Biological Science, Graduate School of Science, Nagoya University, Nagoya, Japan

[3]Department of Mathematics, Pusan National University, Busan, Republic of Korea

[4]Department of Epidemiology and Biostatistics, Indiana University School of Public Health, Bloomington, IN, USA

[5]Shanghai Institute of Infectious Disease and Biosecurity, Fudan University, Shanghai, China

[6]School of Public Health, Fudan University, Key Laboratory of Public Health Safety, Ministry of Education, Shanghai, China

[7]Department of Infectious Diseases, Huashan Hospital, Fudan University, Shanghai, China

[8]Center for Health Emergencies, Bruno Kessler Foundation, Trento, Italy

[9]Laboratory for the Modeling of Biological and Socio-technical Systems, Northeastern University, Boston, Massachusetts, USA

[10]Lee Kong Chian School of Medicine, Nanyang Technological University, Singapore, Singapore

*Corresponding author: majelli@iu.edu

# Abstract

Infectious disease spread is a multi-scale process composed of within-host (biological) and between-host (social) drivers and disentangling them from each other is a central challenge in epidemiology. Here, we introduce VIBES, a multi-scale modeling framework that explicitly integrates viral dynamics based on patient-level data with population-level transmission on a data-driven network of social contacts. Using SARS-CoV-2 as a case study, we analyze three emergent epidemic properties, namely the generation time, serial interval, and pre-symptomatic transmission. First, we established a purely biological baseline, thus independent of the reproduction number ($R$), from the within-host model, estimating a generation time of 6.3 days for symptomatic individuals and 43.1% presymptomatic transmission. Then, using the full model incorporating social contacts, we found a shorter generation time (5.4 days at $R$=3.0) and an increase in pre-symptomatic transmission (52.8% at $R$=3.0), disentangling the impact of social drivers from a purely biological baseline. We further show that as pathogen transmissibility increases ($R$ from 1.3 to 6), competition among infectious individuals shortens the generation time and serial interval by up to 21% and 13%, respectively. Conversely, a social intervention, like isolation, increases the proportion of pre-symptomatic transmission by about 30%. Our framework also estimates metrics that are challenging to obtain empirically, such as the generation time for asymptomatic individuals (5.6 days; 95%CI: 5.1-6.0 at $R$=1.3). Our findings establish multi-scale modeling as a powerful tool for mechanistically quantifying how pathogen biology and human social behavior shape epidemic dynamics as well as for assessing public health interventions.

Introduction

Understanding and predicting the course of an epidemic is a fundamental scientific challenge, primarily because infectious disease spread is a multi-scale process. At one level is the within-host (biological) scale, where the replication dynamics of a pathogen and the host's immune response jointly determine an individual's outcomes (e.g., severity, infectiousness profile over the course of their infection)[1,2]. At another level is the between-hosts (social) scale, where human behavior, contact patterns, and social structures create the opportunities for transmission to occur[3,4]. These scales are inextricably interconnected[5], and a central objective of modern epidemiology is to disentangle their respective contributions in order to improve our ability to understand and predict epidemic trajectories and to design effective control measures.

Recent studies have leveraged viral load data collected from SARS-CoV-2 patients throughout the course of infection to develop mathematical models focused on viral replication and elimination within a single host. These models have been instrumental in characterizing SARS-CoV-2 viral load in individual cases of infection, providing critical insights into infectiousness profiles[6], guidelines for the implementation of screenings and isolation strategies[7–9], and the design of trials for antiviral drugs[10]. However, they lack information about the contact network of infected individuals, which is key for the transmission dynamics of respiratory pathogens within a population[3,11]. This limits the capacity of within-host models to evaluate the effectiveness of a wide range of intervention options considered by public health authorities. Conversely, between-hosts mathematical models, focus on pathogen transmission between infected individuals and their susceptible contacts[12]. Such models have been widely used to improve our epidemiological understanding of spreading patterns and to evaluate the effectiveness of interventions[13,14]. However, they often lack the resolution to account for individual-level heterogeneities, such as differences in the clinical course of infection and viral shedding patterns, which shape epidemiological trends and the effectiveness of interventions[15]. Such simplification makes it difficult to mechanistically connect what happens inside the host to the transmission patterns observed at the population level, thereby limiting our ability to disentangle the impact of biological and social drivers of the epidemic dynamics.

A recent literature review[5] on linking within-host and between-host scales identified several studies that have contributed to the theoretical foundations of a multi-scale framework. However, few incorporated empirical virological data, and none integrated social contact data. To address this gap, we developed VIBES (Viral dynamics Individual-Based Epidemic Simulator), a multi-scale framework that explicitly integrates within-host viral dynamics based on viral dynamics data with between-host transmission dynamics using a social network derived from data on in-person human contact patterns. The within-host component simulates viral kinetics using differential equations whereas the between-hosts component uses an agent-based model that stochastically simulates transmission among a synthetic population structured to incorporate main transmission settings for respiratory pathogens such as households, schools, and workplaces. This allows critical epidemiological characteristics to emerge naturally from the simulation's core biological mechanisms and social interactions rather than being assumed upfront.

To illustrate the model's capability to disentangle these components, we apply VIBES to the transmission dynamics of the ancestral SARS-CoV-2 lineage. We focus on three epidemiological metrics of high public health relevance, all sensitive to the interaction between viral kinetics and social contact (i) the generation time, the interval between infection events in an infector–infectee pair; (ii) the serial interval, the interval between symptom onsets in successive case generations; and (iii) the proportion of pre-symptomatic transmission, the fraction of secondary infections occurring during the pre-symptomatic phase of the primary case. By simulating epidemic spread under various conditions, our study quantifies how pathogen-specific traits and population-level social dynamics independently shape the emergent properties of epidemic dynamics.

## Results

### The Multi-Scale Framework

VIBES is a multi-scale, agent-based model of SARS-CoV-2 transmission that combines: the viral replication of the pathogen within a single individual (within-host model) and the transmission of the pathogen between individuals (between-hosts model).

- The within-host model is based on a system of differential equations simulating viral replication and elimination within each infected agent where the joint posterior distributions of model parameters were estimated based on data from 210 SARS-CoV-2 infected individuals[16]. The within-host model simulates individual trajectories of the viral load over the course of the infection.
- The between-hosts model leverages a synthetic population of agents that is statistically equivalent to the population of Indiana, USA, which is structured to reflect social contact patterns data within households, schools, workplaces, and the broader community[17]. Transmission of the pathogen between infected and susceptible synthetic individuals occurs in four social settings: households, schools, workplaces, and the community.

When an agent is infected in the between-hosts model, an independent run of the within-hosts model is performed to calculate the viral load trajectory for that agent. This trajectory determines the infectiousness profile over time and the epidemiological status of the agent in the between-hosts model (i.e., latent, infectious, recovered, etc.), effectively integrating the biological and social drivers. Details of the models are reported in the Methods section and the Supplementary Material.

VIBES records the entire transmission chain (who infected whom) between the simulated agents and allows for the simulation of individual-level interventions. In the model, the pathogen's transmissibility is calibrated to achieve a target reproduction number, (i.e., the number of secondary infections caused by a primary infector, $R$). Because the agent-based framework does not yield a closed-form expression for $R$, this calibration is performed by adjusting transmissibility parameters and numerically estimating $R$ from simulated epidemic realizations, as described in the Methods section. In this study, we modeled isolation of symptomatic individuals in their place of residence, as an example of an intervention that has traditionally been widely adopted at the onset of epidemic outbreaks. A schematic representation of VIBES is shown in Figure 1; details on the model are reported in the Methods and Supplementary Material.

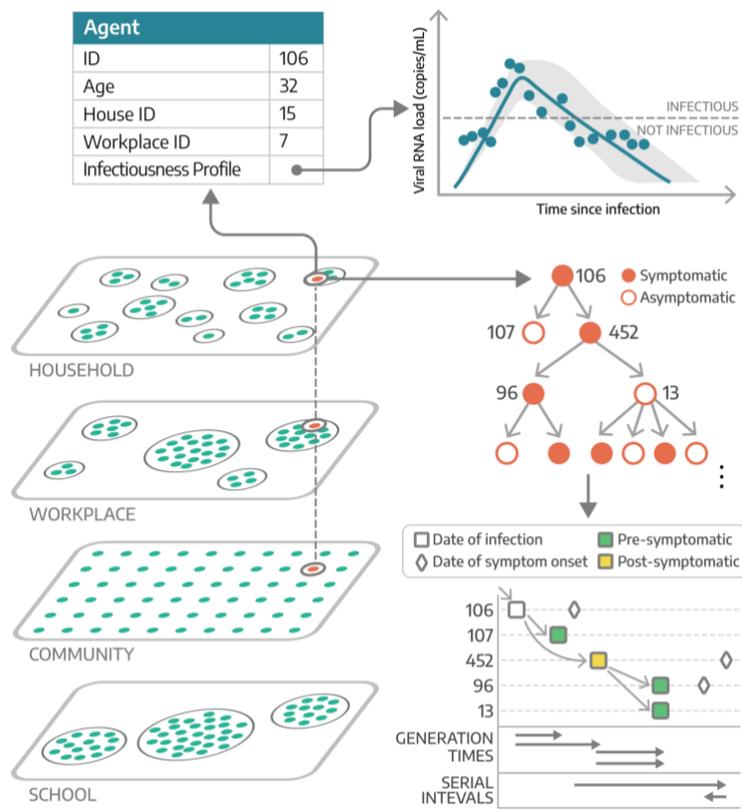

**Figure 1. Conceptual representation of VIBES.** We use a synthetic population of 500,000 agents, statistically equivalent to the population of Indiana, USA[17]. Transmission between individuals occurs in four settings: Households, schools, workplaces and the community. The within-host model simulates the viral replication and elimination within each host, providing an individual viral load trajectory that determines the infectiousness profile and epidemiological status of each infected individual. By tracking individual infections, we directly calculate the following transmission metrics: (i) generation time, defined as the time between the infection event of an infector and of their infectee(s); (ii) serial interval, the time between the symptom onset of an infector and of their infectee(s); and (iii) pre-symptomatic transmission, defined for symptomatic infectors as the proportion of secondary infections that occur before the symptom onset of the infector.

To validate VIBES, we consider three key epidemiological metrics:

- The generation time, which is defined as the time interval between the infection event of an infector and their infectee(s). The generation time is a key determinant of the overall speed of epidemic growth with a shorter value resulting in a faster epidemic growth. In turns, this demands a rapid deployment of public health response to contain and/or mitigate an outbreak[39].
- The serial interval, which represents the time elapsed between the onset of symptoms in an infector and their infectee(s). The serial interval is important to inform contact tracing operation as well as isolation and quarantine guidelines. These types of non-pharmacological interventions are cornerstones of the public health response in the early phase of an outbreak when pharmacological options such as vaccines and antiviral treatments are not available[40].
- The pre-symptomatic transmission, which quantifies, for a symptomatic infector, the proportion of secondary infections that occur before the infector develops symptoms, relative to the total infections caused during their entire infectious period. Pre-symptomatic transmission poses a significant challenge for outbreak control, as it allows the pathogen to spread silently within a population. This highlights the need for measures that target transmission from individuals who are infectious but not yet showing symptoms such as widespread testing, mask-wearing, and social distancing[41].

These metrics were selected for this study because they are largely affected by pathogen transmissibility, individual infectiousness (which in VIBES is determined by the viral dynamics model), and human contact patterns (which may be affected by public health interventions). For example, as transmissibility increases, the generation time and serial interval become shorter due to an increase in competition between infectious individuals to infect susceptible individuals[42]. This is particularly evident in social contexts with a limited number of individuals (e.g., households). Pre-symptomatic transmission increases as transmissibility increases due to the same mechanism. Regarding public health interventions, for example, when home isolation of identified infected individuals is in place, isolation shortens both the generation time and serial interval while it increases pre-symptomatic transmission.

It is important to stress that VIBES does not explicitly include the concept of the generation time, serial interval, and pre-symptomatic transmission (i.e., they are not model parameters). The combination of the within-host and between-hosts scales allows these metrics to emerge from the epidemic dynamics and be measured from the analysis of the transmission chain of simulated epidemics.

## Model Validation

We validated model estimates against the ones obtained from the empirical studies (see Methods). For the generation time, empirical studies estimated a mean generation time for the ancestral SARS-CoV-2 lineages of 5.1 days (95%CI: 3.4-6.5) for reproduction numbers between 1.3 and 2.8[18–24]. Leveraging the transmission chains simulated by VIBES and using the same range for the reproduction number found in the analyzed empirical studies, we estimated a mean symptomatic generation time (i.e., the infector is symptomatic) of 5.7 days (95%CI: 5.5-6.0) when no isolation policies are implemented and 5.0 days (95%CI: 4.8-5.2) when all symptomatic individuals are isolated at home after symptom onset (Fig. 2A). See Supplementary Material for the detail on the estimation of the uncertainty.

For the serial interval, empirical studies estimated a mean value of 5.4 days (95%CI: 3.6-7.4) for reproduction numbers ranging from 1.3 to 3.3[22,25–28,30,32,33,35–38]. From VIBES, we estimated a mean serial interval of 5.8 days (95%CI: 5.5-6.0) when no isolation policies are implemented and 4.6 days (95%CI: 4.3-4.9) when all symptomatic individuals are isolated at home after symptom onset (Fig. 2B).

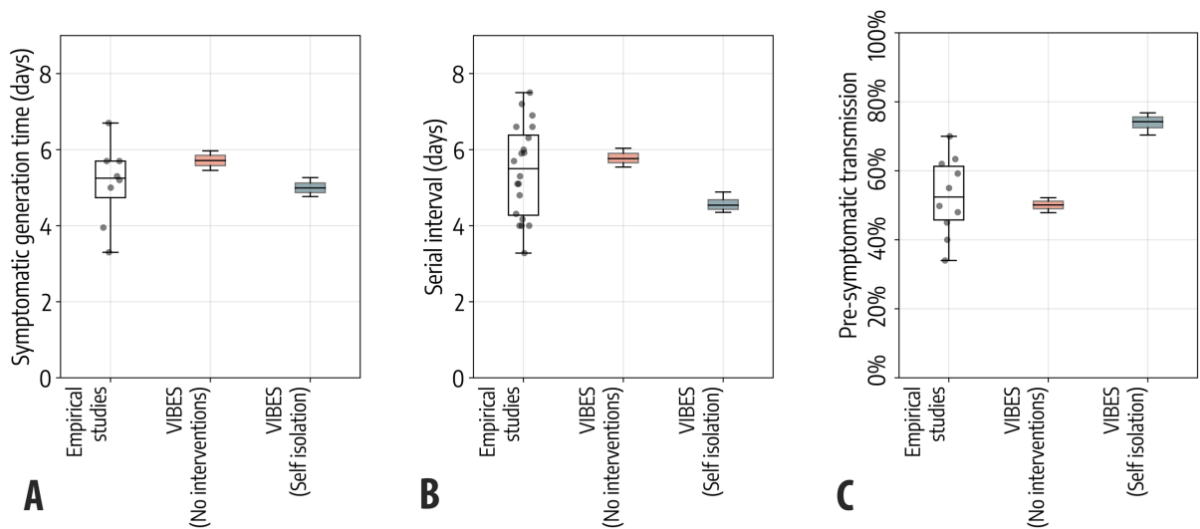

**Figure 2. Estimated epidemiological metrics. A** Boxplot of the distribution of the generation time for symptomatic individuals as estimated in independent empirical epidemiological studies[18–24] and as estimated by VIBES in the absence or presence of isolation of symptomatic individuals. In the identified epidemiological studies, the reproduction number was estimated in the range 1.3-2.8. The same range was used for the estimates provided by VIBES. The box represents the 50% interquantile range, the horizontal line represents the median and the whiskers indicate the farthest data point within 1.5 times the inter-quartile range from the box. Dots represent point estimates from each independent epidemiological study. **B** As A, but for the serial interval. Empirical estimates were taken from refs. [19,22,25–38], where the reproduction was estimated in the range 1.3-3.3. The same range was used for the estimates provided by VIBES. **C** As A, but for the proportion of pre-symptomatic transmission. Empirical estimates were taken from refs. [18,19,21–24,30,32], where the reproduction was estimated in the range 1.3-2.8. The same range was used for the estimates provided by VIBES.

For pre-symptomatic transmission, empirical studies estimated a mean of 52.6% (95%CI: 35.4-68.5) for reproduction numbers ranging from 1.3 to 2.8[18,19,21–24,30,32]. From VIBES, we estimated a mean of 50.1% (95%CI: 48.0-52.1) when no isolation policies are implemented and 74.9% (95%CI: 70.6-76.7) when all symptomatic individuals are isolated at home after symptom onset (Fig. 2C).

The larger variability observed in the empirical estimates as compared to our model-based estimates is associated with the heterogeneities between studies in terms of pathogen transmissibility (reproduction number), implemented public health interventions, study location (social contacts and demographics), etc., with single studies even reporting multiple estimates for different time periods and locations[43,44]. Our model allows us to adjust for these heterogeneities by simulating transmission in a controlled setting (location, reproduction number, and intervention) to assess the effect of transmissibility and isolation on the generation time, serial interval, and pre-symptomatic transmission.

### Disentangling the Biological and Social Components of Transmission

VIBES offers a level of control that allows the independent assessment of how social behavior and biological processes shape key epidemiological metrics. While the within-host model can estimate these metrics based solely on the biological interaction between pathogen and host, VIBES integrates this information with the effects of pathogen transmissibility, contact patterns, and interventions such as isolation. Details on how we estimated epidemiological metrics from the within-host model and with VIBES for a range of $R$ values can be found in the Methods section.

To disentangle the biological from social effects, first, we ran the within-host model alone, which considers only biological factors. We obtained estimates for the mean generation time of 6.3 days for symptomatic and 6.5 for asymptomatic individuals, irrespective of the reproduction number (Fig. 3A,B). These estimates are derived using normalized infectiousness profiles obtained from the sampled viral load trajectories, as detailed in the Methods section. Then, we ran VIBES, which considers both biological and social factors, obtaining lower estimates of the generation time. For a reproduction number of 3.0, VIBES estimates a generation time of 5.4 days (95%CI: 5.3-5.4) for symptomatic and 5.7 days (95%CI: 5.6-5.7) for asymptomatic individuals, a 14.3% and 12.3% decrease from the within-host model estimates, respectively (Fig. 3A,B).

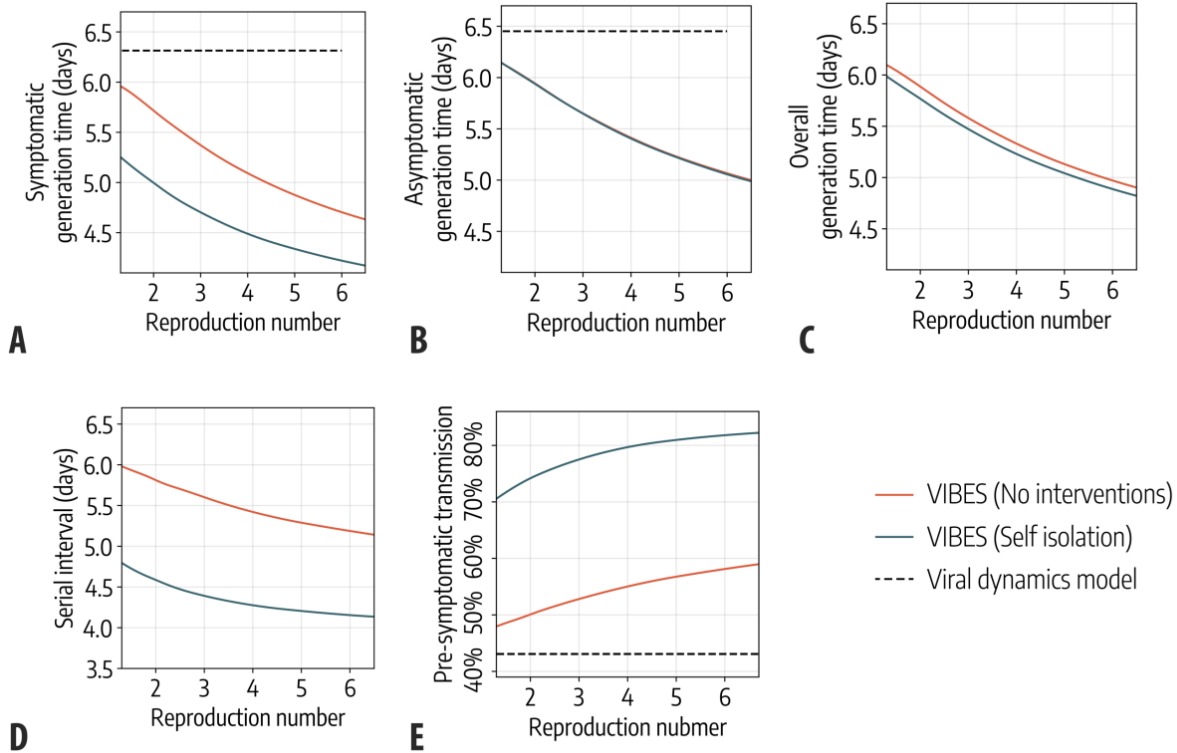

**Figure 3. Effect of transmissibility and isolation on epidemiological metrics. A** Average generation time for symptomatic individuals estimated by VIBES in the presence and absence of isolation of symptomatic individuals as a function of the reproduction number. The dashed line represents the estimated average generation time using the viral dynamics model only, which does not depend on the value of the reproduction number. **B** As A but for the generation time for asymptomatic individuals. **C** As A but for the generation time for all infected individuals. **D** As A but for the average serial interval. **E** As A but for the proportion of pre-symptomatic transmission.

Estimates from VIBES also show a decreasing trend of the generation time with increasing transmissibility. Assuming no isolation policies, VIBES shows a 21.7% decrease in the mean generation time of symptomatic individuals with a reproduction number ranging from 1.3 to 6.0, going from 6.0 days (95%CI: 5.9-6.0) to 4.7 days (95%CI: 4.7-4.7) for reproduction numbers of 1.3 and 6, respectively (Fig. 3A). For asymptomatic individuals, VIBES shows a 16.4% decrease in generation time in the same range for the reproduction number, going from 6.1 days (95%CI: 6.1-6.2) to 5.1 days (95%CI: 5.1-5.1) (Fig. 3B). These lead to an estimate of the overall generation time that lies in between the two (Fig. 3C).

When considering home isolation for symptomatic individuals, we estimated a 20.7% decrease in the generation time for symptomatic individuals as the reproduction number increases, going from 5.3 days (95%CI: 5.2-5.3) to 4.2 days (95%CI: 4.2-4.2) from a reproduction number of 1.3 to 6.0 (Fig. 3A). Since we assumed no isolation of asymptomatic individuals, the estimated generation time for asymptomatic infectors is not affected by the intervention (Fig. 3B) and only a marginal effect is found for the overall estimate (Fig. 3C).

The serial interval showed a similar decreasing trend for increasing transmissibility. Using VIBES and assuming the absence of isolation policies, we estimated that the serial interval decreases from 6.0 days (IQR: 5.9-6.1) to 5.2 days (95%CI: 5.2-5.2) as the reproduction number increases from 1.3 to 6, respectively, corresponding to a 13.3% decrease (Fig. 3D). When considering 100% isolation of symptomatic individuals, we estimated a decrease in the serial interval from 4.8 days (IQR: 4.7-4.9) to 4.2 days (IQR: 4.1-4.2) as the reproduction number increased from 1.3 to 6.0, respectively, corresponding to a 12.5% decrease. When isolation is considered, the estimated serial interval is about 20% shorter than that estimated in the absence of isolation, regardless of the value of the reproduction number (Fig. 3D).

Again, to disentangle the biological and social factors, first we run the within-host viral dynamics model alone finding a 43.1% fraction of pre-symptomatic transmission. Using VIBES, assuming a reproduction number $R = 3.0$ and in the absence of isolation, we estimated a pre-symptomatic transmission of 52.8% (95%CI: 52.4-53.3), which is 22.5% higher than the estimate that considers only biological drivers.

As transmissibility increases, more pre-symptomatic transmission is observed. Using VIBES and having no isolation policies in place, we estimated a 17.3% increase in pre-symptomatic transmission, ranging from 48.0% when $R = 1.3$ to 58.1% when $R = 6$ (Fig. 3E). Isolation of symptomatic individuals has a marked effect on pre-symptomatic transmission that increases to 70.5% when $R = 1.3$ and to 81.8% when $R = 6$ (Fig. 3E).

We conducted a set of sensitivity analyses to assess the robustness of our results to different modeling assumptions. We found that our results are only marginally affected by changes in the probability of developing symptoms and the infectiousness of symptomatic individuals relative to asymptomatic ones (Fig. S3 and S4 in the Supplementary Material). Small quantitative differences were observed when infectiousness was homogeneous between individuals (i.e., no superspreading), although the qualitative trends were confirmed (Fig. S5 in the Supplementary Material).

## Discussion

We validated the VIBES model against empirical epidemiological data for the ancestral lineage of SARS-CoV-2, showing good agreement between our model-based estimates of key epidemiological metrics and those obtained from traditional field epidemiological studies. By combining individual-level viral load data with detailed synthetic contact networks, VIBES provides novel insights into epidemiological characteristics that are challenging to measure in the field. Specifically, we estimated the generation time for asymptomatic individuals to be 5.6 days (95%CI: 5.1-6.0 days) when $R = 1.3$. Our study also characterizes the associations between pathogen transmissibility, interventions, and the key epidemiological metrics. Specifically, we found that with increasing pathogen transmissibility (from $R = 1.3$ to $6$), the generation time and serial interval decreased by 21% and 13%, respectively, which is compatible with findings from field studies showing a progressive shortening of these two indicators as new more transmissible variants emerged[45]. Moreover, the contribution of pre-symptomatic transmission increased by 17% ($R$ from $1.3$ to $6$). Isolation of symptomatic individuals shortened both generation time and serial interval by about 2% and 20%, respectively, and increased the proportion of pre-symptomatic transmission by about 30%. These findings show the complex interplay between pathogen transmissibility, individual-level interventions, and key epidemiological metrics that shape outbreak dynamics.

A strength of VIBES is its ability to estimate critical epidemiological metrics from the early stages of an outbreak, drawing on data sources that are potentially available in real time. From the one side, VIBES relies on individual-level viral load data, which can be approximated from PCR testing that often becomes available early in an outbreak, as shown by the deployment of PCR diagnostics within two weeks since the admission of the first COVID-19 patients in a Wuhan hospital[25,46,47]. From the other side, VIBES relies on contact pattern data that can be derived from pre-existing sources, including publicly available micro- and macro-level datasets, social contact studies, and synthetic populations representing specific locations[17,48–53]. Alternatively, contact patterns can be approximated using mobility data collected in real time [54–57]. It is important to emphasize that our approach is not intended to replace traditional epidemiological studies, but to complement them by providing an additional analytical tool for understanding the spread of novel pathogens. Indeed, estimating metrics like the generation time (especially for asymptomatic individuals) and the fraction of pre-symptomatic transmission poses several challenges in traditional field studies. Empirical estimates also exhibit larger variability due to heterogeneities in study populations, social behaviors, and implemented interventions resulting in wide confidence interval and making results of different studies often hard to reconcile[45,58–61]. VIBES, however, can derive these estimates in a controlled setting from first principles (i.e., they are emerging properties of the model).

The epidemiological metrics used in this study (i.e., generation time, serial interval, and proportion of pre-symptomatic transmission) are crucial for understanding the epidemiology of novel pathogens and designing effective control measures. For instance, the generation time dictates the pace of the epidemic, providing information on the timeline of deployment of public health interventions[39]. The serial interval provides insights into the design of symptom-based interventions such as case isolation[40]. The fraction of pre-symptomatic transmission provides information about the need for interventions that go beyond symptom-based approaches. However, it is important to stress that other epidemiological metrics (e.g., reproduction number, infection fatality risk, infection hospitalization risk) are also crucial for preparedness and response planning[62,63]. This emphasizes again the importance of using a range of complementary approaches during an epidemic outbreak to provide timely and reliable insights on its epidemiology, transmission dynamics, and public health impact.

It is important to acknowledge the limitations of our study. First, our analysis uses the ancestral lineage of SARS-CoV-2 as a case study. While the identified qualitative trends may be common to other pathogens, the quantitative estimates obtained in this study could hardly be generalized to other SARS-CoV-2 variants or other pathogens. Second, the within-host model of viral dynamics is directly taken from our previous study[16]. While this can be seen as a strength of this study, as it relies on a well-established and validated approach, other viral dynamics models have been introduced in the literature considering more complex viral dynamics (e.g., considering an eclipse phase slowing down viral growth[64] or an innate immune response – interferons[33]). A new version of VIBES could benefit from more refined within-host models. Third, VIBES, in its current form, relies on a simplified representation of human behavior and does not account for potential changes in social contact patterns or adherence to interventions over time. These can affect both qualitative trends and quantitative estimates of epidemiological metrics. Moreover, our analysis considers only one type of intervention: isolation of symptomatic individuals. While isolation is a crucial control measure, it is just one component in a wide array of public health interventions considered by policy makers. For instance, antiviral treatment can reduce viral load of treated patients, which may in turn lower their transmission potential. Finally, our analysis focuses only on a single geographic location, using a synthetic population representative of Indiana, USA. While we expect the identified patterns to hold true for other locations, quantitative estimates may be different.

In conclusion, we introduced a novel multi-scale modeling framework, VIBES, that bridges the gap between within-host viral dynamics and population-level transmission, disentangling the influence of biological and social drivers of transmission. By providing a mechanistic understanding of key epidemiological metrics and their interplay with transmissibility and interventions, VIBES offers a valuable tool for informing public health strategies, particularly in the critical early stages of an emerging infectious disease outbreak.

## Methods

### Within-host model

For this study, we developed the viral dynamics model presented in Jeong et al.[16]. Briefly, the model consists of a set of two ordinary differential equations accounting for the fraction of uninfected target cells and the number of viruses per unit of sample specimens (i.e., viral load in copies/ml). A nonlinear mixed effects model was used to fit empirical viral load data retrieved from 210 patients infected with the ancestral SARS-CoV-2 lineage (109 symptomatic, and 101 asymptomatic) to estimate the posterior distribution of three free model parameters. The procedure was independently performed for symptomatic and asymptomatic patient data, resulting in different parameter distributions. The infectiousness profile was assumed to be proportional to the logarithm of the viral load (copies/ml) using a threshold of $10^{5.0}$ copies/ml, under which an individual is not able to transmit the virus[66,67]. For symptomatic infections, the incubation period (time from infection to symptom onset) was estimated from the viral load to match the empirical distribution estimated in Hu et al.[19], for which an average of 6.4 days (IQR: 3.2-8.8) was found. Details on the viral dynamics model can be found in our previous study[16] and in the Supplementary Material.

### VIBES

In our model, transmission between infectious and susceptible individuals takes place among a synthetic population of agents that can be represented as a multiplex layer network with four layers: household, school, workplace, and general community. The synthetic population has been taken from our previous work[17], and it comprises 200,000 households (approximately 500,000 individuals) representative of the Indiana (USA) population. Briefly, each individual in the synthetic population is assigned to a household and the general community. Based on employment and school enrollment rates by age, each individual is assigned to a workplace and/or a school (if any) in the respective layers. Homogeneous mixing is assumed within each household, workplace, school, and in the general community.

The transmission model features six epidemiological statuses to which each individual is assigned at each step of the simulation: susceptible, latent, infectious pre-symptomatic, infectious symptomatic, infectious asymptomatic and removed. Susceptible individuals can contract the infection through contact with an infected individual. Upon infection, we use an age-dependent probability[68] to sample whether the individual will develop symptoms or remain asymptomatic. Then a viral load profile is generated by the viral dynamics model, where the parameters are sampled from the distribution fitted for symptomatic or asymptomatic patients separately. At the first simulation step after the simulated viral load exceeds the infectiousness threshold, the (latent) individual becomes able to transmit the virus; when the threshold is crossed once again in the declining phase of the infection, the individual is no longer able to transmit the virus and is considered to be removed.

The transmission probability $p_{i \to j}(t)$ that infectious individual $i$ infects susceptible individual $j$ that is in contact with $i$ in setting $l \in \{\text{household, workplace, school, community}\}$ at time $t$ is defined as:

$$p_{i \to j}(t) = \alpha \cdot \frac{\beta_l}{N_{i,l}} \cdot g_i \cdot u_i(t - \tau_i) \cdot \Delta t,$$

where:
- $\beta_l$ is the setting-specific transmission risks, which were estimated to match the proportions of infections by social setting reported in Liu et al.[13];
- $\alpha$ is a scaling rate that is set to obtain the desired value of the reproduction number;
- $N_{i,l}$ is the number of individuals in setting $l$ to which individuals $i$ and $j$ belong;
- $\tau_i$ is the time when individual $i$ was infected;
- $u_i(t - \tau_i)$ is the infectiousness profile of individual $i$ at step $t - \tau_i$, determined by the viral dynamics model (see Supplementary Material for details);
- $g_i$ is a coefficient determining the overall infectiousness of individual $i$, which is sampled from a gamma distribution with shape=0.235 and scale=4.26 as estimated in Sun et al.[22];
- $\Delta t = 0.25$ days is the time step of the simulation.

We also implemented a 100% isolation policy for symptomatic individuals, where they isolate in their place of residence since the day of symptom onset until the end of the infectious period. When an individual is isolated, they continue to have contacts with their household members, while they do not have any contacts with individuals in other social settings.

Details on the between-hosts transmission model are reported in the Supplementary Material.

## Estimation of epidemiological metrics using VIBES

During the simulation, we track the entire transmission chains and produce a simulated line list of patients that contains information about the infector, infection date, and date of symptom onset. From this line list, we calculated epidemiologically relevant metrics. Specifically:

- The reproduction number was calculated as:

$$R = \frac{1}{M_W} \sum_{i \,|\, t_i \in W} r_i$$

where $W$ is a subset of all time steps of a simulation where the infection incidence shows an exponential growth (Details in the Supplementary Material), $r_i$ is the number of secondary infections caused by individual $i$, $t_i$ is the time in which individual $i$ was infected, $M_W = \sum_{i \,|\, t_i \in W} 1$ is the number of individuals that were infected in time window $W$, and $i \,|\, t_i \in W$ selects only individuals that were infected in time window $W$.

- The serial interval $S_I$ is calculated as:

$$S_I = \frac{1}{N_F} \sum_{e \in F} (\tau_e - \sigma_e)$$

where $F$ is the set of infector-infectee pairs in which both individuals develop symptoms, $\tau_e$ is the time at which the infected individual developed symptoms, $\sigma_e$ is the time at which the infector individual developed symptoms and $N_F$ is the number of elements in $F$.

- The overall generation time $T_G$ was calculated as:

$$T_g = \frac{1}{N_E} \sum_{e \in E} (t_e - s_e)$$

where $E$ is the set of all infector-infectee pairs, $t_e$ is the time at which the infection occurred, $s_e$ is the time at which the infector was infected, and $N_E$ is the number of infections during the simulation. Similarly, we defined the symptomatic generation time and asymptomatic generation time by limiting the set $E$ to include only symptomatic or asymptomatic infectors, respectively.

- The fraction of pre-symptomatic transmission $P_T$ was calculated:

$$P_T = \frac{1}{N_G} \sum_{e \in G} J(t_e < \sigma_e)$$

where $G$ is the set of infector-infectee pairs in which the infector is symptomatic, $N_G$ is the number of elements in set $G$, and $J$ is an indicator variable that is 1 if the condition is true, 0 otherwise.

### Estimation of epidemiological metrics using the viral dynamics model

We used the within-host component of VIBES to estimate epidemiologically relevant metrics directly from the simulated viral trajectories. Specifically:

- The generation time was estimated as:

$$T_g = \frac{1}{N_P} \sum_{k \in P} \sum_{t=0}^{\infty} u_k(t) \cdot t$$

  where $P$ is the set of viral load profiles of either symptomatic or asymptomatic patients, $N_P = 10{,}000$ is the number of samples and $u_k(t)$ is the normalized infectiousness from the sampled viral trajectory $k$ at time $t$ measured after infection, so that $\sum_{t=0}^{\infty} u_k(t) = 1, \forall\, k \in P$.

- The proportion of pre-symptomatic transmission was estimated as:

$$P_T = \frac{1}{N_P} \sum_{k \in P} \sum_{t=0}^{\tau_k - 1} u_k(t)$$

  where again $P$ is the ensemble of viral load profiles for symptomatic individuals, and $\tau_k$ is the incubation period for sampled viral load profile $k$.

### Literature estimates of epidemiological metrics

For validation and comparison with the VIBES model, we collected estimates of the generation time, serial interval, and pre-symptomatic transmission from a systematic literature review[45]. Among the papers cited in the literature review, we selected only those that: i) provide at least one estimate of the generation time, serial interval, or fraction of pre-symptomatic transmission for the ancestral SARS-CoV-2 lineage, and ii) provide an estimate of the reproduction number that is greater than 1, ensuring that the metrics were estimated in the growing phase of an outbreak.

### Funding

PCV, ML, AGK, AV, and MA acknowledge the support of the cooperative agreement CDC-RFA-FT-23-0069 from the CDC's Center for Forecasting and Outbreak Analytics. PCV, ML, AGK, AV, and MA acknowledge support from the Cooperative Agreement no. NU38OT000297 of the Centers for Disease Control and Prevention (CDC) and the Council of State and Territorial Epidemiologists (CSTE). Its contents are solely the responsibility of the authors and do not necessarily represent the official views of the CDC and CSTE. YDJ was supported by the National Research Foundation of Korea (NRF) grant funded by the Korean government (MSIT) (grant number: RS-2024-00345478). KE was supported by a Singapore Ministry of Education startup grant (LKCMedicine-SUG, #022487-00001). HY was supported by the Key Program of the National Natural Science Foundation of China (grant number 82130093 to H.Y.) and the Shanghai Municipal Science and Technology Major Project (grant number ZD2021CY001). SM was funded by the NextGeneration EU-MUR PNRR Extended Partnership initiative on Emerging Infectious Diseases, Project no. PE00000007, INF-ACT. This research was supported in part by Lilly Endowment, Inc., through its support for the Indiana University Pervasive Technology Institute.

## Competing interests

HY has received research funding from Sanofi Pasteur, Shenzhen Sanofi Pasteur Biological Products Co., Ltd, Shanghai Roche Pharmaceutical Company, and SINOVAC Biotech Ltd. None of the research funding is related to this work.

# VIBES: A Multi-Scale Modeling Approach Integrating Within-Host and Between-Hosts Dynamics in Epidemics

# SUPPLEMENTARY MATERIAL

Paulo C. Ventura[1], Yong Dam Jeong[2,3], Maria Litvinova[4], Allisandra G. Kummer[1], Shingo Iwami[2], Hongjie Yu[5,6,7], Stefano Merler[8], Alessandro Vespignani[9], Keisuke Ejima[10], Marco Ajelli[1,*]

[1]Laboratory for Computational Epidemiology and Public Health, Department of Epidemiology and Biostatistics, Indiana University School of Public Health, Bloomington, IN, USA

[2]Interdisciplinary Biology Laboratory (iBLab), Division of Biological Science, Graduate School of Science, Nagoya University, Nagoya, Japan

[3]Department of Mathematics, Pusan National University, Busan, Republic of Korea

[4]Department of Epidemiology and Biostatistics, Indiana University School of Public Health, Bloomington, IN, USA

[5]Shanghai Institute of Infectious Disease and Biosecurity, Fudan University, Shanghai, China

[6]School of Public Health, Fudan University, Key Laboratory of Public Health Safety, Ministry of Education, Shanghai, China

[7]Department of Infectious Diseases, Huashan Hospital, Fudan University, Shanghai, China

[8]Center for Health Emergencies, Bruno Kessler Foundation, Trento, Italy

[9]Laboratory for the Modeling of Biological and Socio-technical Systems, Northeastern University, Boston, Massachusetts, USA

[10]Lee Kong Chian School of Medicine, Nanyang Technological University, Singapore, Singapore

*Corresponding author: majelli@iu.edu

# Table of Content

## Methods

### Viral dynamics model

Longitudinal viral load data was taken from Jeong et al.[1], which includes samples for 210 patients who met the criteria (109 symptomatic and 101 asymptomatic patients). We used a simple viral dynamics model to describe the evolution of the viral load over the course of infection. This model has been successfully employed for infectious diseases that cause infection, including SARS-CoV-2[2,3]:

$$\frac{df(t)}{dt} = -\beta f(t)V(t), \qquad (1)$$

$$\frac{dV(t)}{dt} = \gamma f(t)V(t) - \delta V(t). \qquad (2)$$

The first variable $f(t)$ is the ratio between the numbers of uninfected target cells at time $t$ and 0. The second variable $V(t)$ is the number of viruses per unit of sample specimens (copies/ml) at time $t$. The parameters in the model, $\beta$, $\gamma$, and $\delta$ are the rate of virus infection, the maximum rate of viral replication, and the virus clearance rate of infected cells, respectively. The time $t$ is measured in days, with $t = 0$ representing the day of infection. Following the previous study from Goyal et al. [4], the viral load at the day of infection was set as $V(0) = 10^{-2}$ (copies/mL), and $f(0) = 1$ by definition. Under reasonable parameter setting, the time course change of viral load $V(t)$ typically presents a bell-shaped curve (see Fig. S1); the viral load exponentially increases after infection, hits the peak, and starts declining once a substantial number of cells were infected, which is empirically observed in longitudinal viral load data that cause acute infections.

### Fitting procedure

We employed a nonlinear mixed effects model[5] for fitting the viral dynamics model to the longitudinal viral load data. The fitting was performed on the software MONOLIX 2019R2[6]. The nonlinear mixed-effect model incorporates two separated effects on each parameter: fixed effect and random effect. The fixed effect captures population-level dynamics (thus it is the same for all individuals), while the random effect explains the individual variability in the dynamics. Specifically, the parameter for a person $i$, $\theta_i (= \theta \times e^{\eta_i})$ is a product of $\theta$ (fixed effect) and $\eta_i$ (random effect). The random effect follows a normal distribution with mean 0 and standard deviation $\Omega$: $N(0, \Omega)$. The population parameters $\theta$ (i.e., fixed effect) were estimated using Stochastic Approximation Expectation Maximization (SAEM). Individual parameters $\theta_i$ (i.e., combination of fixed and random effects) were estimated using Markov Chain Monte Carlo (MCMC). The SAEM and MCMC algorithms were employed because of the complex landscape of the likelihood functions, which demand high-dimensional parameter space exploration. The parameter estimation process for fixed effect was separated into two steps: one is the exploring phase and the second is the smoothing phase. In brief, the parameter region with maximum likelihood was explored and identified in the first step, then the identified parameter region was smoothed to identify the maximum likelihood estimate in the second step. Individual parameters (i.e., fixed + random effect parameters) were estimated as empirical bayes estimates (EBEs). Figure S1 shows the estimated viral load dynamics for randomly selected 8 symptomatic and 8 asymptomatic patients, respectively. In the simulation, longitudinal viral load data were generated by running the viral dynamics model with a parameter set randomly sampled from the estimated posterior parameter distributions.

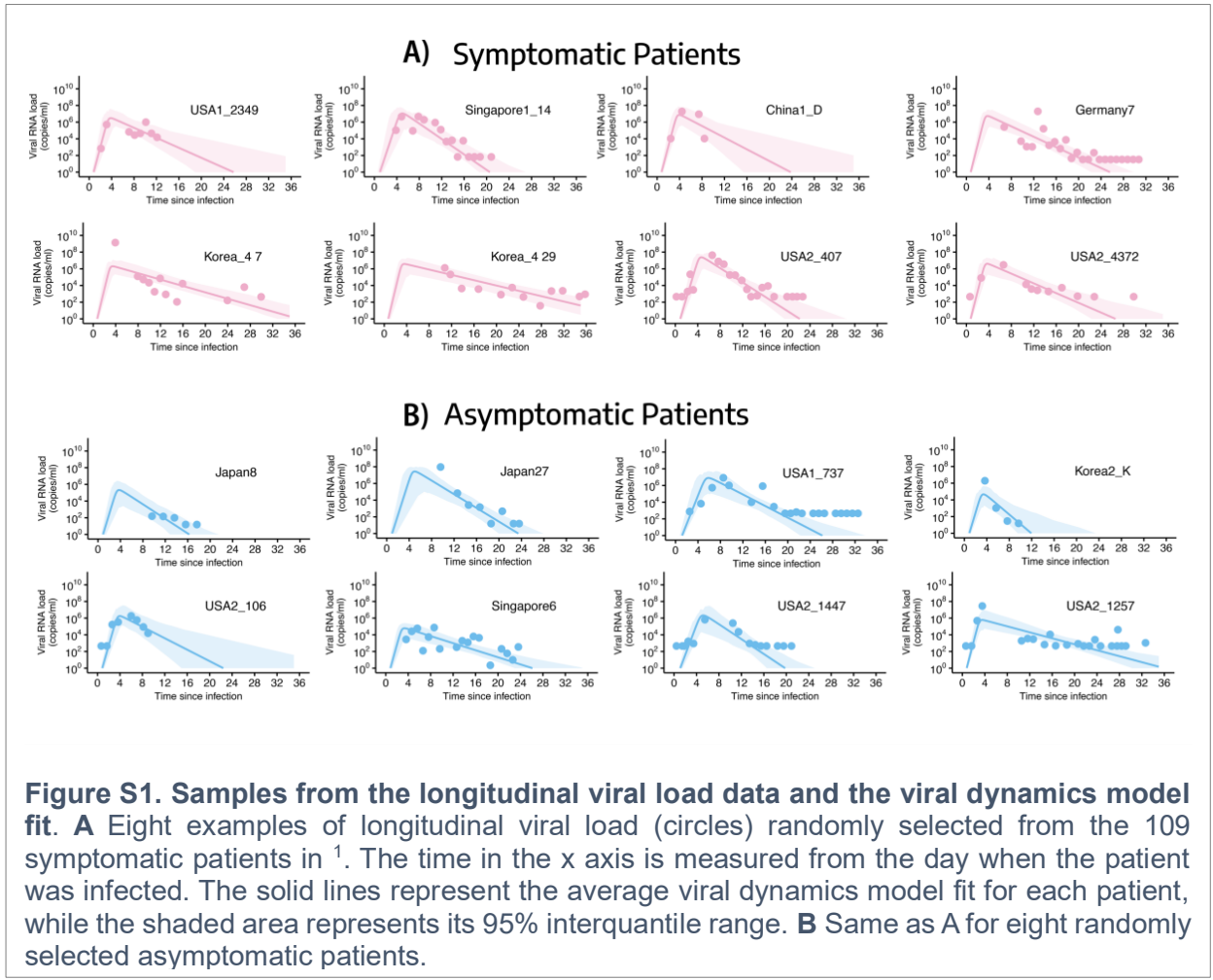

**Figure S1. Samples from the longitudinal viral load data and the viral dynamics model fit**. **A** Eight examples of longitudinal viral load (circles) randomly selected from the 109 symptomatic patients in [1]. The time in the x axis is measured from the day when the patient was infected. The solid lines represent the average viral dynamics model fit for each patient, while the shaded area represents its 95% interquantile range. **B** Same as A for eight randomly selected asymptomatic patients.

## Conversion of viral load to infectiousness (and infectiousness threshold)

The infectiousness was defined as the per contact probability of transmission in this study. As observed in multiple epidemiological and experimental studies, the infectiousness becomes negligible when the viral load is below a threshold value (i.e., the infectiousness threshold)[7]. In this study, we assumed that an individual is infectious when the viral load is above the threshold, and non-infectious otherwise. The infectiousness was assumed as logarithmically proportional to the viral load during the infectious period. We set the infectiousness threshold as $10^{5.0}$ copies/ml following epidemiological and experimental studies[8,9]. Specifically, the normalized infectiousness $u_i(t)$ for an individual $i$ at time $t$ after their infection was computed from viral load using the following equation:

$$u_i(t) = \begin{cases} A_i \cdot \log(V_i(t)), & t \in [t_{i,0}, t_{i,1}], \\ 0, & \text{otherwise,} \end{cases} \quad (3)$$

where the viral load is above the infectiousness threshold from $t_{i,0}$ to $t_{i,1}$ (i.e., $V_i(t_{i,0}) = V_i(t_{i,1}) = 10^{5.0}$ copies/ml). The normalization constant $A_i$ is defined as:

$$A_i = 1 \bigg/ \sum_{t=t_{i,0}}^{t_{i,1}} \log V_i(t) . \quad (4)$$

The length of the infectious period is defined as $t_{i,1} - t_{i,0}$, assumed to be contiguous because the viral dynamics is unimodal under reasonable parameter settings.

### Simulation of incubation period using viral dynamics model

We assumed that the symptom starts during when the viral load is above $(100 - m)\%$ of the peak viral load. Specifically, the incubation period for an infected individual $i$ was computed by the following process:

1) A longitudinal viral load data sample was generated by running the viral dynamics model with a parameter set randomly sampled from the estimated posterior parameter distributions. The peak viral load for an individual $i$ is denoted by $V_i^*$.
2) We computed the time interval when the viral load is above $(100 - m)\%$ of the peak viral load, denoted by $[t_{i,0}^m, t_{i,1}^m]$. Note that $V_i(t_{i,0}^m) = V_i(t_{i,1}^m) = V_i^*(1 - m/100)$.
3) The incubation period for individual $i$ ($T_i$) was then calculated as the sum of $t_{i,0}^m$ and a random number from an exponential distribution $X$: $X \sim Exp(\lambda)$: $T_i = t_{i,0}^m + X$.
4) Step 3) was repeated until the incubation period was in the interval $[t_{i,0}^m, t_{i,1}^m]$ and $[t_{i,0}, t_{i,1}]$: $T_i \in [t_{i,0}^m, t_{i,1}^m] \cap [t_{i,0}, t_{i,1}]$.

The two parameters, $m$ and $\lambda$ were determined so that the generated incubation period distribution matches the incubation period estimated by Hu et al. in an epidemiological study[10]: a Weibull distribution with mean 6.4 days (median: 5.7 and IQR: 3.2-8.8). Specifically, $m$ and $\lambda$ were computed to minimize the following sum squared error (SSE):

$$SSE(m, \lambda) = (Mean - 6.4)^2 + (Median - 5.7)^2 + (Q25\% - 3.2)^2 + (Q75\% - 8.8)^2,$$

where $Mean, Median, Q25\%$ and $Q75\%$ are respectively the mean, median, 25% quantile and 75% quantile of generated distribution. By following this procedure, we obtained an incubation period distribution with median 5.5 days and 50% interquantile range of 4.0–7.5 days.

### Synthetic population

To model the contacts relevant for the transmission of SARS-CoV-2, we used the high-resolution multiplex network synthesized from real-world sociodemographic data that was developed in Mistry et al.[11]. This network consists of four layers representing contact settings: household, school, workplace, and community. The network features 200,000 households with about 500,000 individuals, corresponding to a representative subsample of the population of Indiana, USA. Each individual is assigned to a household, represented as a cluster in the household layer, and to the community, which assembles all individuals into a single cluster. Moreover, based on age-specific employment and school attendance probability, each individual can be assigned to a workplace (a cluster in the workplace layer) and a school (a cluster in the school layer). The community layer contains all individuals in a single cluster. Within each layer, clusters are connected such that all individuals within each cluster are connected, and no connections between individuals of different clusters exist.

### VIBES model for SARS-CoV-2 transmission

In our individual-based model, SARS-CoV-2 spreads through contacts in the described network. At each time step during the simulation, an epidemiological status is associated to each individual: susceptible, latent, infectious pre-symptomatic, infectious symptomatic, infectious asymptomatic, and removed. Susceptible individuals can acquire the infection from infectious individuals that share a cluster with them. Once infected, a susceptible individual is assigned to the latent status, meaning that they carry the pathogen but cannot infect other individuals. Based on the estimated age-dependent probability of developing symptoms estimated by Poletti et al.[12] (see Table S1), we determine whether an infected individual will develop symptoms during the course of infection. If symptoms are to develop, we sample the viral dynamic parameters from the posterior distribution that was fitted to symptomatic patient data. Otherwise, we sample parameters from the distribution fitted to asymptomatic patient data. We then obtain a viral load trajectory by simulating the within-hosts model with the selected parameters.

Over the course of an infection, the viral load trajectory determines the progression of an individual through epidemiologic statuses. Specifically, when the viral load exceeds a threshold of $10^5$ copies/mL[7], the individual becomes able to transmit the pathogen. At this point, individuals flagged to develop symptoms are assigned to the infectious pre-symptomatic status, while individuals who will not develop symptoms are assigned to the infectious asymptomatic status. When the time of symptom onset is reached, determined by the incubation period, infectious pre-symptomatic individuals are assigned to the symptomatic status. Finally, symptomatic and asymptomatic individuals are assigned to the removed status once their viral load goes below the same threshold of $10^5$ copies/mL, at which point they are unable to infect other individuals and remain so until the end of the simulation.

The within-host viral dynamics also determine the infectiousness of an individual during their infectious period. The probability $p_{i \to j}(t)$ that infectious individual $i$ infects susceptible individual $j$ that is in contact with $i$ in setting $l \in \{\text{household}, \text{workplace}, \text{school}, \text{community}\}$ at time $t$ is defined as:

$$p_{i \to j}(t) = \alpha \cdot \frac{\beta_l}{N_{i,l}} \cdot g_i \cdot u_i(t - \tau_i) \cdot \Delta t, \tag{5}$$

Where:

- $\alpha$ is the overall transmission rate;
- $\beta_l$ is the layer-specific relative transmission risk;
- $N_{i,l}$ is the number of individuals in the given cluster of layer $l$;
- $u_i(t - \tau_i)$ is the infectiousness of individual $i$ at step $t - \tau_i$, where $\tau_i$ represents the time of infection of individual $i$;
- $g_i$ is a gamma-distributed random variable determining the overall infectiousness of individual $i$, with shape 0.235 and scale 4.26 to obtain the distribution of the number of secondary infections reported in Sun et al[13].
- $\Delta t = 0.25$ days is the duration of a single time step in the simulation.

The rate $\alpha$ and the layer-specific transmission risks $\beta_l$, were set such that the reproduction number in the exponential phase was $R = 1.3$ and the proportions of infections in each layer matched those estimated by social settings, as reported in Liu et al.[14] (see Table S1). Each simulation was initialized with 30 latent individuals, with all other individuals susceptible. All model parameters are reported in Table S1.

**Table S1.** Model parameters. List of model parameters and values used in the main analysis with no interventions.

| Parameter description | Parameter value |
|---|---|
| Study population | Indiana, USA |
| Number of individuals (population size) | 502,697 |
| Number of households | 200,000 |
| Length of a time step | 0.25 days |
| Initial number of latent individuals | 30 |
| Setting-specific transmission risks (fraction of transmission by setting) [#]: | |
|    Households | 1.0 {35%} |
|    Schools | 0.418 {15%} |
|    Workplaces | 0.307 {15%} |
|    Community | 0.304 {35%} |
| Individual transmissibility coefficient[@]: | |
|    Gamma distribution's shape | 0.24 |
|    Gamma distribution's scale | 4.13 |
| Infectiousness of symptomatic individuals relative to asymptomatic ones | 1 |
| Probability of developing symptoms by age group[$]: | |
|    0-19 years old | 18.1% |
|    20-39 years old | 22.4% |
|    40-59 years old | 30.5% |
|    60-79 years old | 35.5% |
|    80+ years old | 54.6% |

[#] Transmission risks and fraction of transmission by setting refer to the scenario R=1.3, as estimated in Hu et al.[10] and Liu et al.[14]
[@] Parameters set to obtain the distribution of the number of secondary infections reported in Sun et al.[13]
[$] Reference: Poletti et al.[12]

## Time window for the estimation of the reproduction number

From the synthetic line list of infections generated with VIBES, we can directly calculate the reproduction number of a simulation over time and for the entire simulation. The reproduction number is defined as the average number of secondary infections that a typical infected individual generates over the course of their infection, characterizing the overall infectiousness of the pathogen during one simulation. However, since the reproduction number decreases over time as immunity builds, we calculate the reproduction number only during early stages of each epidemic. The result allows us to compare different simulations in terms of the infectiousness.

We use thresholds in the cumulative incidence of infections to determine the window of calculation of the reproduction number. Given one simulation, the start of the calculation window $w_0$ is defined as the first time step $t$ in which the cumulative incidence of infections is greater than or equals to $c_0 = 1000$ individuals. The end of the calculation window $w_1$ is defined as the first time step $t$ in which the cumulative incidence of infections is greater than or equals to $c_1 = 7000$ individuals. The threshold $c_1$ for the end of the calculation window also defines the occurrence of an outbreak: If the infection incidence never reaches $c_1$ by the end of the simulation, we consider that it did not produce an outbreak, and the reproduction number is not calculated for that simulation. For this work, we only considered simulations that produce an outbreak for calculating the transmission pattern metrics.

## Estimation of the uncertainty

All the presented results are based on a high-resolution grid of values of pathogen transmissibility. Specifically, we predefined a grid of values for the overall transmissibility ($\alpha$) to simulated different values of the reproduction number. We then simulated 50 independent stochastic executions of the model for each $\alpha$ in the pre-selected grid. For each value of $\alpha$, we calculated the mean values of $R$, $T_g$, $S_I$ and $P_T$ from the ensemble of independent executions. Finally, to estimate the uncertainty on each epidemiological metric, we used an exact quadratic spline interpolation with $R$ as the independent variable and the desired parameter ($T_g$, $S_I$ or $P_T$) as dependent variable, such that $R$ lies within the target range. Confidence intervals represent quantiles of this interpolated ensemble.

For the estimation of the epidemiological metrics derived from the literature (i.e., those used to validate model estimates), we used an unweighted average of the estimates reported in each study. The uncertainty provided on these estimates represent the interquartile range of the means.

## Supplementary Results

Simulations using the VIBES model typically show a single-peak outbreak pattern infection incidence over time. With a fixed set of parameters, we performed $n = 50$ simulations and calculated the number of new infections at each time step. Since the length of the time step is $\Delta t = 0.25$ days, we aggregated the number of infections at each day to obtain a daily time series of the infection incidence. We then shifted the trajectory of each simulation such that time $t = 0$ represents the first time step in which the cumulative infection incidence was greater than 1 infection per 1000 individuals. With all simulations aligned in such way, we calculated the daily average incidence, as well as the daily 95% interquantile range of the incidence. We obtained simulations for different values of the reproduction number by changing the scaling rate $\alpha$. The results indicate that simulations display a typical single-peak outbreak pattern, which reproduces the patterns obtained with traditional between-hosts simulations when immunity wane is not considered (Fig. S2). This shows that our model is able to reproduce patterns of other models that have been successfully employed to simulate infectious disease outbreaks.

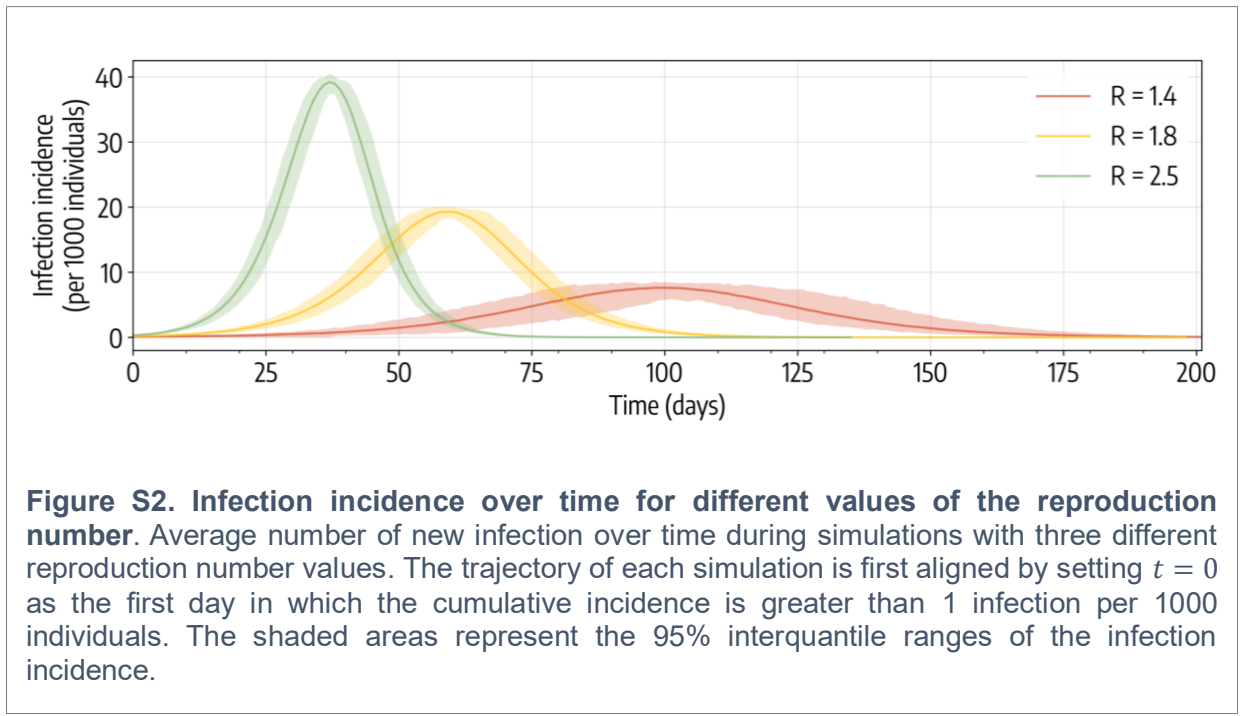

**Figure S2. Infection incidence over time for different values of the reproduction number.** Average number of new infection over time during simulations with three different reproduction number values. The trajectory of each simulation is first aligned by setting $t = 0$ as the first day in which the cumulative incidence is greater than 1 infection per 1000 individuals. The shaded areas represent the 95% interquantile ranges of the infection incidence.

## Sensitivity analysis: Probability of symptom development

We tested different assumptions for the probability that each infected individual develops symptoms. This was achieved by multiplying the age-dependent probabilities described in Table S1 by a uniform factor of 1.5 (50% higher probability of symptom development) and 0.5 (50% lower probability of symptom development). The estimates for the generation time of symptomatic and asymptomatic infectors were robust to changes in this assumption. The overall generation time shifts towards lower values when the symptom development probability is higher. This happens because symptomatic individuals, on average, have lower generation time than asymptomatic ones, and therefore increasing the ratio of symptomatic infections over asymptomatic ones biases the generation time towards lower values. Similarly, the overall generation time shifts towards higher values when the probability of symptom development is lower. No significant effects of changing the symptom development probability were observed into the serial interval and the pre-symptomatic transmission (Fig. S3).

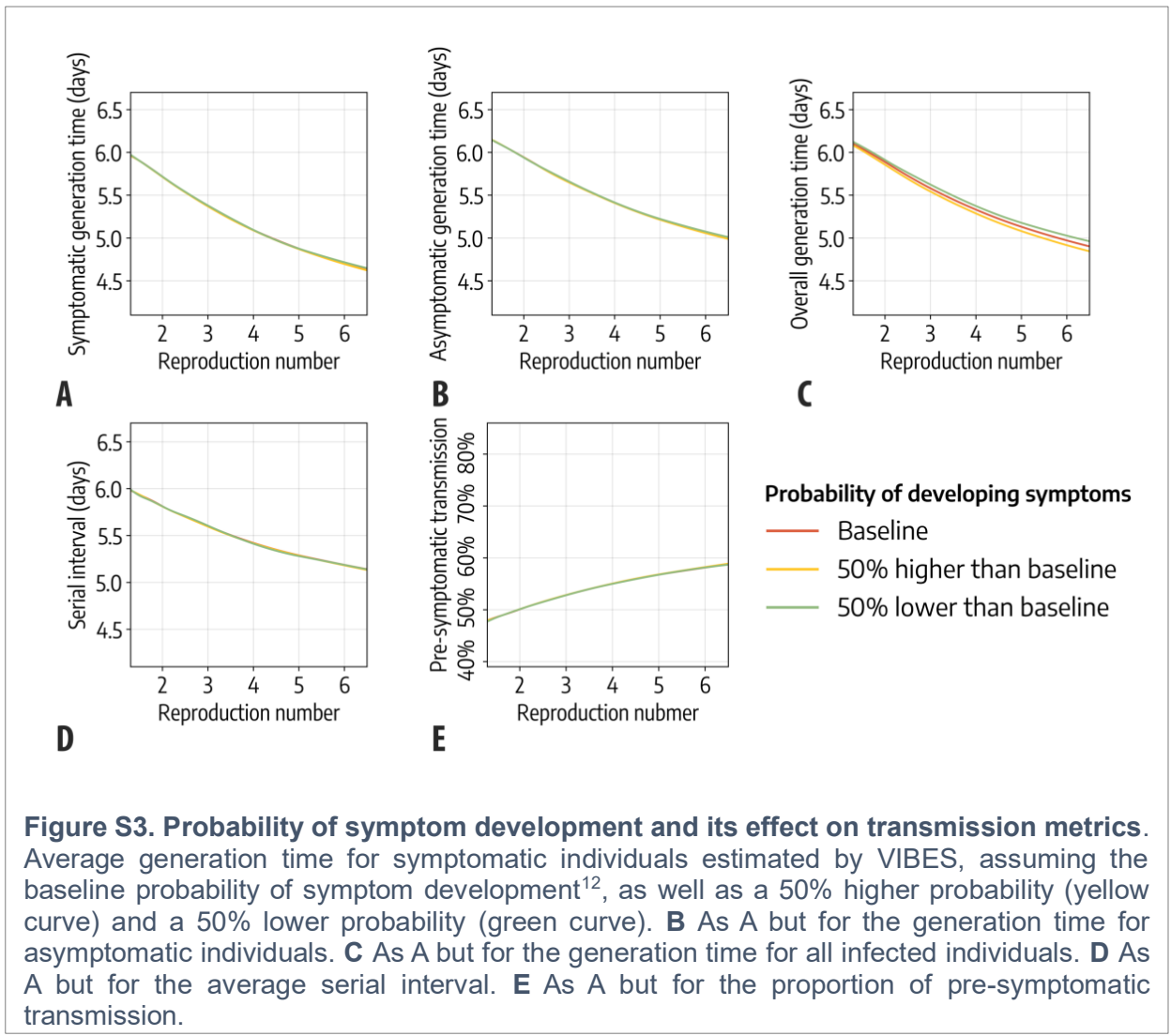

**Figure S3. Probability of symptom development and its effect on transmission metrics**. Average generation time for symptomatic individuals estimated by VIBES, assuming the baseline probability of symptom development[12], as well as a 50% higher probability (yellow curve) and a 50% lower probability (green curve). **B** As A but for the generation time for asymptomatic individuals. **C** As A but for the generation time for all infected individuals. **D** As A but for the average serial interval. **E** As A but for the proportion of pre-symptomatic transmission.

## Sensitivity analysis: Infectiousness of symptomatic individuals relative to asymptomatic ones

Our results are also robust to variations in the infectiousness of symptomatic individuals when compared to asymptomatic ones. To study that, we increased the infectiousness created by symptomatic individuals by a factor of 1.5 (50% higher than baseline) compared to that of asymptomatic ones. This led to only marginal changes in the generation time, serial interval and fraction of pre-symptomatic transmission, effectively showing no impact of such change in the studied transmission patterns. Similarly, no impacts were observed when the infectiousness of symptomatic individuals was multiplied by a factor of 0.75 (25% lower than baseline) (Fig. S4).

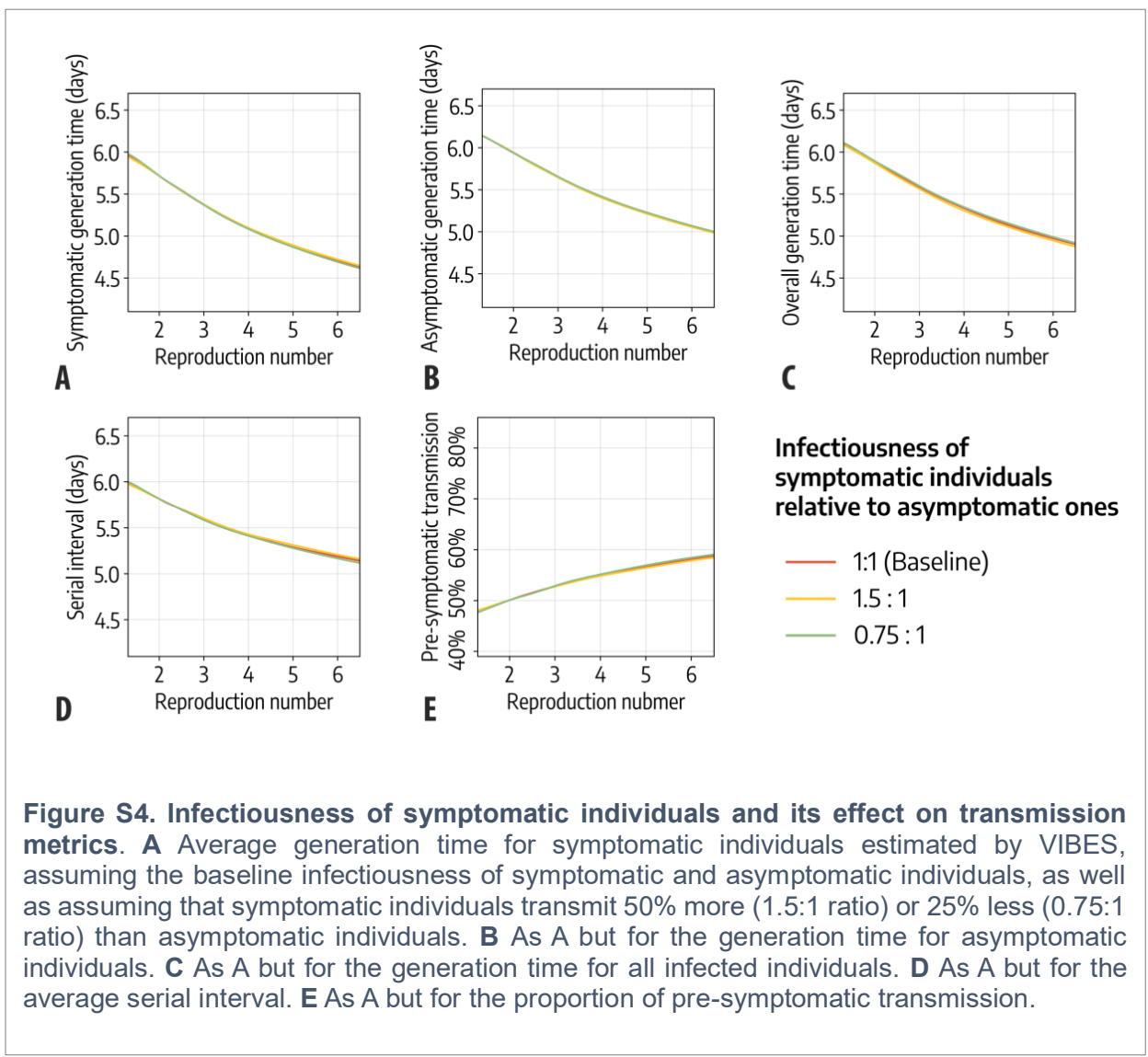

**Figure S4. Infectiousness of symptomatic individuals and its effect on transmission metrics**. **A** Average generation time for symptomatic individuals estimated by VIBES, assuming the baseline infectiousness of symptomatic and asymptomatic individuals, as well as assuming that symptomatic individuals transmit 50% more (1.5:1 ratio) or 25% less (0.75:1 ratio) than asymptomatic individuals. **B** As A but for the generation time for asymptomatic individuals. **C** As A but for the generation time for all infected individuals. **D** As A but for the average serial interval. **E** As A but for the proportion of pre-symptomatic transmission.

## Sensitivity analysis: Distribution of the numbers of secondary infections

Transmission of SARS-CoV-2 displays significant overdispersion[13], meaning that the distribution of the number of secondary infections has a negative-binomial shape, with superspreading events likely. In the baseline model, we applied a gamma-shaped distributed coefficient to the infectiousness of each individual which, convolved with the transmission process, creates a negative binomial distribution similar to the one observed in the field. As a sensitivity analysis, we removed this individual coefficient, which results in a Poisson-shaped distribution of secondary cases. This caused small changes in the estimate epidemiologic metrics estimates, while the overall trend remained the same. The Poisson-distributed infectiousness implied shorter generation times for both symptomatic and asymptomatic individuals, as well as shorter serial intervals and larger pre-symptomatic transmission (Fig. S5).

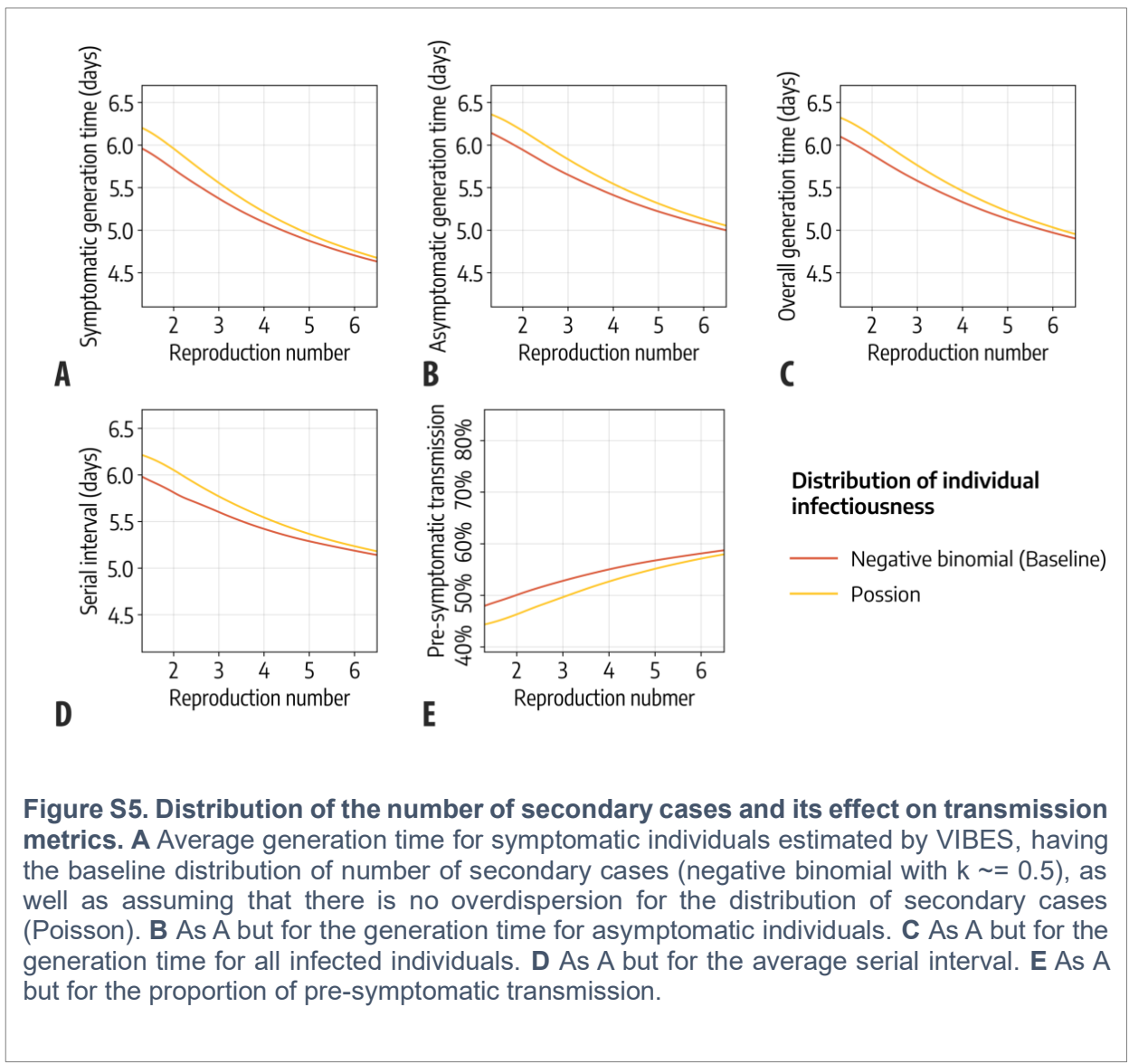

**Figure S5. Distribution of the number of secondary cases and its effect on transmission metrics. A** Average generation time for symptomatic individuals estimated by VIBES, having the baseline distribution of number of secondary cases (negative binomial with k ~= 0.5), as well as assuming that there is no overdispersion for the distribution of secondary cases (Poisson). **B** As A but for the generation time for asymptomatic individuals. **C** As A but for the generation time for all infected individuals. **D** As A but for the average serial interval. **E** As A but for the proportion of pre-symptomatic transmission.